\documentclass{sistedes}
\usepackage{graphicx}
\usepackage[style=ieee,backend=biber]{biblatex}   
\usepackage{multirow}

\usepackage{arxiv}

\begin{document}

\title{A comparison between traditional and Serverless technologies in a  microservices setting\thanks{Supported by University of Oviedo and DXC Technology}}
\titlerunning{A comparison between traditional and Serverless technologies}
\author{Juan Mera Menéndez \and 
 José Emilio Labra Gayo \and
Enrique Riesgo Canal \and 
Aitor Echevarría Fernández }

\authorrunning{J. Mera et al.}

 \institute{
 University of Oviedo, Asturias, Spain\\
 \and
 DXC Technology\\
}

\maketitle              
\begin{abstract}
Serverless technologies, also known as FaaS (Function as a Service), 
are promoted as solutions that provide 
dynamic scalability, 
speed of development, 
cost-per-consumption model, 
and the ability to focus on the code while taking attention away from the infrastructure that is managed by the vendor. 
A microservices architecture is defined by the interaction and management of the application state by several independent services, each with a well-defined domain.
When implementing software architectures based on microservices, there are several decisions to take about the technologies and the possibility of adopting Serverless. 

In this study, we implement 9 prototypes of the same microservice application using different technologies. 
Some architectural decisions and their impact on the performance and cost of the result obtained are analysed. 
We use Amazon Web Services and start with an application that uses a more traditional deployment environment (Kubernetes) and migration to a serverless architecture is performed by combining and analysing the impact (both cost and performance) of the use of different technologies such as AWS ECS Fargate, AWS Lambda, DynamoDB or DocumentDB. 
		
\end{abstract}
	
\section{Introduction}
	
Cloud computing provides a model that enables application deployment with expected lower costs and greater scaling flexibility than more traditional approaches. 
The Serverless approach \parencite{Baldini2017} promises to allow 
developers to focus on the code, 
forgetting to manage the complexity related to the infrastructure. 
By combining the two previous concepts, a high degree of flexibility and speed of reaction to possible changes, among other characteristics, in a development can be achieved. \\
	
An important difficulty of entry arises when a company must make a transition between traditional legacy software and solutions that embrace these new models. 
The need to migrate from more traditional technologies to cloud-native versions frequently appear. 
Thus, it is necessary to make architectural decisions having the best possible information about the consequences of different alternatives. 

The main goal of this research is to shed some light in this direction, 
 by providing a comparison between different alternatives, 
 perform system migrations to better understand their architectural strengths and weaknesses, 
 compare the results in terms of cost and performance and highlight some critical points and troubles of the migration processes.
All the source code of the different migrations is available in a github repository\footnote{\url{https://github.com/catedradxc/serverlessStudy/}}.
 
In order to do that, we depart with a microservice based application that has been implemented using a traditional approach with independent containers deployed in Kubernetes\footnote{\url{https://kubernetes.io/}} and migrate it to alternative Serverless based technologies.
	
We chose Amazon Web Services (AWS)\footnote{\url{https://aws.amazon.com/}} as a cloud provider because 
it is the most widely used and well-known cloud provider, 
although closely followed by Microsoft Azure and Google Cloud Platform~\cite{Griffiths23}. 
Besides that, we used three deployment technologies as the goal of the migrations: 
a more traditional one such as Kubernetes (we understand Kubernetes as a more traditional solution compared to a serverless approach, with the adjective "traditional" being somewhat relative), 
a fully Serverless one such as AWS Lambda and one between both approaches such as ECS Fargate
To add a persistence tier, we used two different services: AWS DocumentDB and DynamoDB , the first of them is not Serverless at all (The standard DocumentDB instances require one to provision and manage its own server instances, which is why they are not considered Serverless) 
and the second one fully Serverless. 
On the other hand, we chose java with the Spring Framework because it was the language in which the initial system was, each language presents different features \parencite{Jackson2018, Shilkov2020}. 

The main contribution of this paper is to offer a comparison in terms of costs and performance of a microservices based prototype implemented with 9 different technology possibilities that combine traditional with Serverless approaches. We consider that both the source code and the results can be helpful to have a better undestanding of the consequences of adopting some of those technologies. 

The structure of the paper is the following. Section~\ref{sec:InitialApp} describes the architecture of the initial application, section~\ref{Sec:Prototypes} describes the prototypes that have implemented using different approaches, section~\ref{sec:CostAnalysis} compares the costs and section~\ref{sec:Performance} compares the performance. Section~\ref{sec:Discussion} presents a discussion which is followed by a description of the related work in section~\ref{sec:RelatedWork} and conclusions and future work in section~\ref{sec:Conclusion}.
	
\section{Initial application} \label{sec:InitialApp}

We chose a microservices based prototype that was already available as a starting point for our study\footnote{The starting prototype is available at: \url{https://github.com/ewolff/microservice-kubernetes}}. 
 One of the reasons to chose that application was its simplicity 
 with the intention of demonstrating that, even though it is a simple solution, 
 some of the difficulties that would be encountered to a greater extent with a more complex example still arise.\\

 \begin{figure*}[h]
\centering
\includegraphics[width=0.75\textwidth]{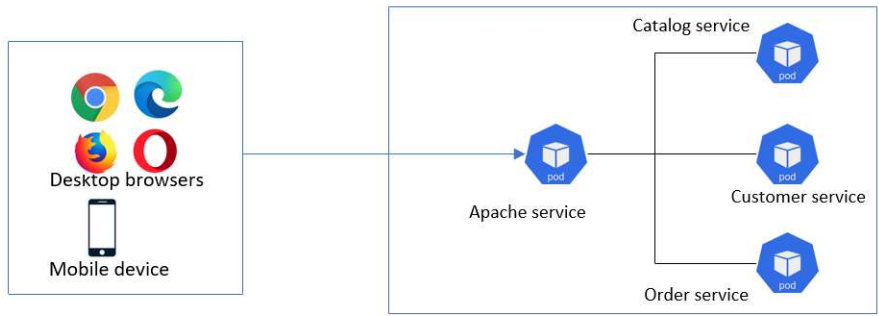}
\caption[Initial application architecture]{Initial application architecture}
\label{fig:initial_architecture}
\end{figure*}

The selected application is presented in~\ref{fig:initial_architecture} and consists of an e-commerce solution, mainly compound of three services with functionality and an extra service to make the orchestration of the three others. 
All the services are self-contained and individual, offering an API through of which the other services send messages. The three principal services will survive along the versions and migrations of the system:
\begin{itemize}
\item Catalogue service: It manages the operations related with the products.
\item Customer service: It manages everything related with customers.
\item Order service: It involves all in relation to orders.
\end{itemize}

The fourth service before mentioned, consists fundamentally in an Apache server to redirect the requests to each corresponding service of the three last. 
This Apache service will be replaced when migrating the application to a more managed technology.

\section{Prototypes and processes} \label{Sec:Prototypes}

Starting from the initial application already outlined, several migration or adaptation processes have been carried out to enable the system to run on other platforms. 
The results of these processes are divided into 3 scenarios, 
each with 3 prototypes: 
the first of them consists of migrating to another platform while maintaining, 
as far as possible, 
the persistence mechanism used by the original version. 
The second one consists of adding a persistence layer to the versions 
of the first scenario, thus analysing the impact of one or another data storage service. 
The third scenario seeks to complete the combination between the technologies 
used and the data storage services. 
Thus, the resulting prototypes are the following:
	
\subsection{AWS EKS}

For this first version, it didn't require much adaptation work to enable the system to be deployed using Elastic Cloud Kubernetes (EKS) as the original application  was already set up for a Kubernetes environment. 
The resulting architecture is the same as the initial application shown in Figure~\ref{fig:initial_architecture}. \\
The process for deployment and configuration includes creating the Kubernetes cluster in EKS with the appropriate permissions, 
and creating the node group where the system will be deployed. 
The chosen EC2 instance type to host the node group is t3.Small, 
as it is the smallest one with enough network interfaces.

\subsection{ECS Fargate}

\begin{figure}[h]
\centering
\includegraphics[width=0.7\linewidth]{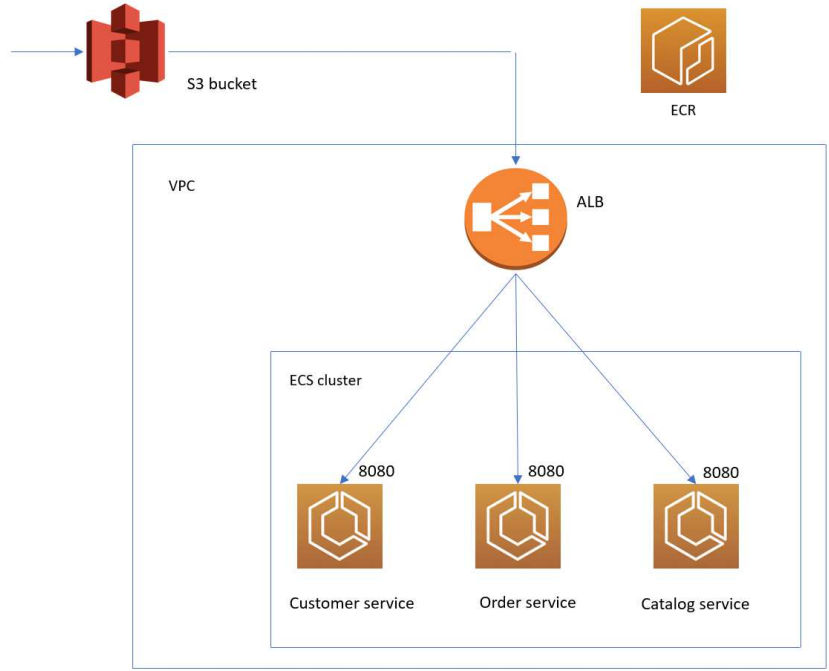}
\caption{ECS fargate architecture}
\label{fig:ECS_Fargate_architecture}
\end{figure}

As a technology that sits in the middle ground between more traditional computing and FaaS, 
it is often chosen as a migration target for legacy software towards a more flexible, scalable, and self-managed technological framework. \\
As that middle ground, many of the components involved in a Kubernetes deployment, for example, are reusable and remain useful in an architecture designed to be deployed on ECS Fargate. 
Although many other elements, such as those related to service orchestration, are replaced by other AWS-specific components like load balancers. \\
In our case, the first service, which consisted of an Apache server, will be replaced by a load balancer with rules configured to perform the same function, redirecting incoming requests to the corresponding service. 
The final architecture of this prototype can be seen in Figure~\ref{fig:ECS_Fargate_architecture} \\
Regarding changes to the code and/or its rewriting, 
there is no need to make any changes starting from the initial version that was containerized and designed for Kubernetes. 
Each image will be deployed on a container within an ECS Fargate task. 
The tasks employed are configured with 0.5 GB of memory and 0.25 vCPU, 
and the containers with a flexible memory limit of 128 MiB. \\
This version has the issue that tasks are not stateless, making it difficult to duplicate for horizontal scaling. 
This problem is solved as soon as data management is delegated to another component.

\subsection{AWS Lambda}

As the final main approach, the use of AWS Lambda service along with API Gateway was chosen as the most widely accepted serverless architecture approach \parencite{Sbarski2018}. 

The back-end will be provided as lambda functions wrapped in an API, and the frontend will be rebuilt as an SPA hosted in an S3 bucket.
As Java with Spring are the technologies chosen, it was necessary to rebuild each service to adapt to the schema (handlers instead of controllers) 
used by AWS Lambda, while keeping the user interface separate. 

This process could be costly and non-trivial depending on the size of the service to be ported and the complexity of its logic. 
In our case, even though the services are relatively small and simple, 
the process required almost a complete rewrite. 

If the services prior to migration were larger in size, 
it would be necessary to split them into functional sections 
of an appropriate size for a lambda function \parencite{Miell2019}. 
This includes dividing the management of the information that these services handle. 

\begin{figure}[h]
\centering
\includegraphics[width=0.7\linewidth]{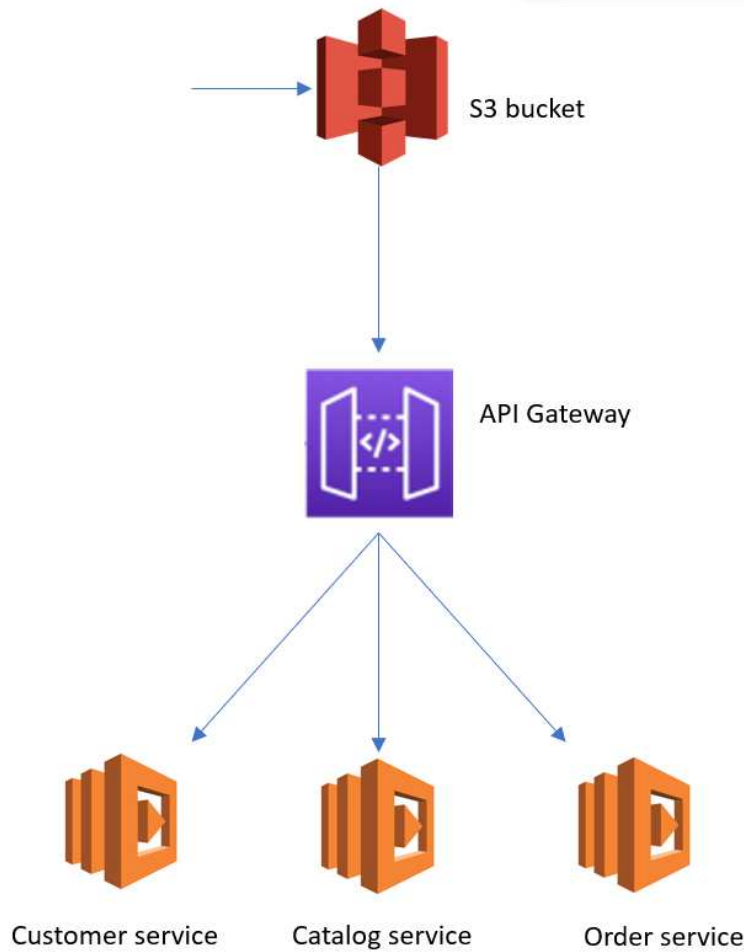}
\caption{AWS lambda architecture}
\label{fig:AWS_lambda_architecture}
\end{figure}

The architecture of this prototype can be seen in Figure~\ref{fig:AWS_lambda_architecture}. The configuration of Lambda functions used is: 512 MB of memory, 512 MB of ephemeral storage, and Java 11 (corretto) as language.

Due to the stateless nature of lambda functions and their scaling approach, the fact that these functions manage data in this first solution makes it not a functional prototype. Again, as soon as the management of the system information is delegated to another component, this problem would be solved.
	
\subsection{AWS EKS with DocumentDB}

A necessary step in the journey to improve the architecture managed in terms of flexibility and scalability, among others, is to add a persistence layer on which to delegate data handling and separate it from logic. 
The services chosen for achieve that challenge are DocumentDB and DynamoDB. For this solution DocumnetDB will be used. 
 
Due to DocumentDB's compatibility with MongoDB, this service is widely used for migrating to cloud environments. 
The changes in the code required to integrate the initial AWS EKS solution with DocumentDB are not too many, and the deployment procedure is the same except for the creation of the DocumentDB cluster. 
As for the DocumentDB configuration, 
version 4.0 will be used with a single instance of type db.t3.medium. 

\begin{figure}[h]
\centering
\includegraphics[width=0.7\linewidth]{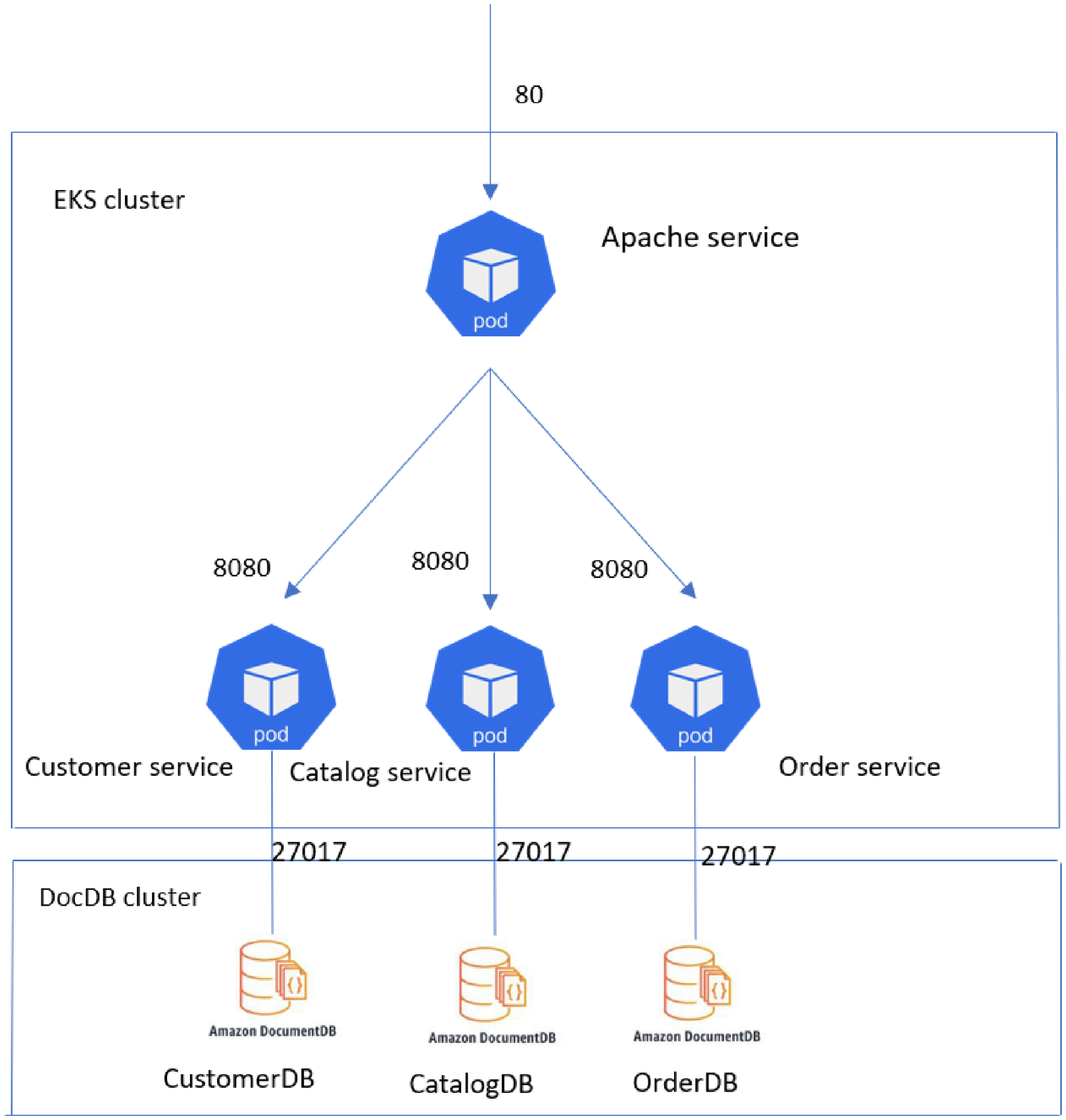}
\caption{EKS + DocumentDB architecture}
\label{fig:EKS_DocumentDB_architecture}
\end{figure}

The resulting architecture can be seen at Figure~\ref{fig:EKS_DocumentDB_architecture}

\subsection{AWS EKS with DynamoDB}

In this case, the prototype combines AWS EKS with DynamoDB, starting from the initial version of AWS EKS prepared for deployment on this service. 
 
To enable the integration of these two services, it is necessary to configure the DynamoDB tables to host the system's data. 
The necessary changes in the code are related to accessing the data from the 3 main services. 
Additionally, it is necessary to provide the appropriate permissions to the EKS worker nodes for proper integration.
 
The AWS EKS configuration used is the one seen before. 
For DynamoDB, the configuration used is the default, with read and write capacity values of 5 RCU and WCU respectively. 
Using provisioned mode with autoscaling enabled.

\subsection{Others}

The last four prototypes combine technologies already discussed in previous ones. They employ similar configurations for all services and without any particularly noteworthy aspect but doing different combinations. 
These four versions are:
	\begin{itemize}
		\item ECS Fargate with DynamoDB
		\item AWS Lambda with DynamoDB
		\item ECS Fargate with DocumentDB
		\item AWS Lambda with DocumentDB
	\end{itemize}
	\vspace{0.2cm}
	
\section{Tests and metrics measured}

In order to validate each of the prototypes developed on one hand, and measure the performance of each of them on the other hand, it is necessary to build a series of tests.
	
\subsection{Validation testing}

To guarantee the validity of the prototypes used for the comparison, an automated  test-suite has been designed using Selenium. 
These tests ensure that all functionalities work from the point of view of a user.

\subsection{Performance testing}

One of the main criteria when comparing architectures is their performance. Thus, in order to determine the performance of each prototype, 
a series of performance and stress tests are designed. 
The objectives of these tests are:

\begin{itemize}
\item To determine the performance of each of the prototypes. This performance will be measured in terms of latency or response time.
\item To determine the load limits supported by each of the prototypes.
\item To compare the different technologies used to create the system versions.
\end{itemize}

As the tests are carried out in a research context, there are no predefined performance standards that the systems must meet. 
The tools chosen were JMeter for load testing and AWS CloudWatch for collecting internal metrics.

All tests are repeated a total of 3 times, with the displayed result being the average value of the three repetitions. 
The performance tests will be carried out with several user loads and a constant ramp-on of 30 minutes. 
For stress testing, the loads will also vary, and the ramp-on will be between 20 and 30 seconds causing sudden load increments.
	
\section{Costs analysis} \label{sec:CostAnalysis}

Another key aspect in choosing architectures is the cost of each one. AWS has services whose costs are primarily calculated in two ways. 
On one hand, we have the traditional cost model of paying for infrastructure rental time in a provisioned resource model. 
On the other hand, some services calculate their price based on consumption, a characteristic model of Serverless computing. 

To enable comparison between technologies that employ the two defined cost models, it is necessary to establish how these costs will be calculated for comparison. Firstly, the currency used will be the euro €. 
For solutions calculated using the traditional hardware rental model, costs will be calculated monthly. 
For versions whose services allow the consumption-based cost model, a constant load of 5 users per second will be assumed, which equates to 13 million requests per month approximately. \\

\begin{table}[h]
\centering
\begin{tabular}{|c|c|}
\hline
\textbf{Prototype} & \textbf{mensual cost (€)} \\ \hline
AWS EKS & 92 \\ \hline
ECS Fargate & 55.8 \\ \hline
AWS Lambda & 17.7 \\ \hline
AWS EKS + DocumentDB & 157.14 \\ \hline
ECS Fargate + DocumentDB & 120.94 \\ \hline
AWS Lambda + DocumentDB & 82.84 \\ \hline
AWS EKS + DynamoDB & 93.53 \\ \hline
ECS Fargate + DynamoDB & 57.33 \\ \hline
AWS Lambda + DynamoDB & 19.23 \\ \hline
\end{tabular}
\caption{Total costs for each prototype}
\label{tabla:costes}
\end{table}

 Starting with solutions based on AWS EKS, this service incurs a cost of 0.1 € per hour, to which the cost of the instance, which amounts to €19 per month, is added. The use of AWS EKS amounts to a total of €92 per month.
 
Continuing with solutions based on ECS Fargate, which use 3 tasks whose cost is €31.09 per month, 1 load balancer whose cost is €19.7 per month, and an S3 bucket whose access costs approximately €5 per month depending on the requests. The use of this architecture amounts to a total cost of €55.8 per month.

As a third approach, solutions based on AWS Lambda with API Gateway, both services with a usage-based cost model. Assuming a load of 13 million requests per month, the total cost will be approximately €17.7 depending on the size of the requests and the execution time of the Lambda functions.

Using DocumentDB with the configuration already mentioned results in a monthly cost of €65.14. On the other hand, using DynamoDB with the mentioned configuration and assuming the monthly load again results in a cost of approximately €1.53.

Thus, summarizing, in table \ref{tabla:costes} you can see the total cost of each of the solutions. 
	
\section{Performance analysis} \label{sec:Performance}

As previously mentioned, the performance of an architecture is always relevant when evaluating different designs. 
The following results were reported by the load and stress tests on the different prototypes built. The data displayed for each version is the results reported in the top 3 load tests that each prototype supported.

\begin{table}[h]
\centering
\begin{tabular}{|c|ccc|ccc|}
\hline
\multirow{2}{*}{Prototype} & \multicolumn{3}{|c|}{\textbf{Performance testing results}} & \multicolumn{3}{|c|}{\textbf{Stress testing results}} \\ 
 & latency(ms) & \%error & \#users & latency(ms) & \%error & \#users \\ \hline
\multirow{3}{*}{AWS EKS} 
  & 	195 & 0 & 1000 		& 220 & 0.74 & 200 \\ 
  & 	161 & 0.05 & 1800 	& 943 & 19.25 & 400 \\ 
  & 	367 & 0.5 & 2500  	& 668 & 23.07 & 500 \\ \hline

\multirow{3}{*}{ECS Fargate} 
  & 468 & 0.5 & 1000 		& 2929 & 0 & 200 \\ 
  &	307 & 0 & 2000 			& 8982 & 4.67 & 400 \\ 
  &	10292 & 48.08 & 5000 	& 16170 & 3.62 & 500 \\ \hline

\multirow{3}{*}{AWS Lambda} 
  & 167 & 0 & 2000 			& 87 & 0 & 400 \\ 
   & 216 & 0 & 5000 		&  77 & 0 & 500 \\ 
  & 196 & 0 & 10000 		& 87 & 0 & 1000 \\ \hline

\multirow{3}{*}{AWS EKS + DocumentDB} 
	&		61 & 0 & 1800 &			63 & 2.83 & 400 \\ 	
	&		60 & 0 & 2500 & 			54 & 21.91 & 500 \\ 
	&		61 & 0 & 10000 &			65 & 34.98 & 1000 \\ \hline

\multirow{3}{*}{ECS Fargate + DocumentDB} 
	&		91 & 0 & 1000 & 			189 & 0 & 500 \\ 
	&		147 & 0 & 2000 & 			852 & 0.02 & 1000 \\ 
	&		119 & 0 & 5000 &			6058 & 1.44 & 2500 \\ \hline

\multirow{3}{*}{AWS Lambda + DocumentDB} 
 &			323 & 0 & 2000 & 			9189 & 0 & 500 \\ 
 &			339 & 0 & 5000 & 			12469 & 0 & 1000 \\ 
 &			261 & 0 & 10000 & 			537 & 62.26 & 2000 \\ \hline

\multirow{3}{*}{AWS EKS + DynamoDB} 
 &		55 & 0 & 1800 & 			56 & 0 & 400 \\ 
 &		55 & 0 & 2500 & 			56 & 0 & 500 \\ 
 &		227 & 1.97 & 10000 &		59 & 27.31 & 1000 \\ \hline

\multirow{3}{*}{ECS Fargate + DynamoDB} 
 &			88 & 0 & 2000  & 			3787 & 0 & 500 \\ 
 &			88 & 0 & 5000  & 			4482 & 3.45 & 1000 \\ 
 &			15872 & 46.81 & 10000 &		1183 & 47.5 & 2500 \\ \hline
\multirow{3}{*}{AWS Lambda + DynamoDB} 
 &			219 & 0 & 1000 & 			1950 & 0 & 400 \\ 
 &			440 & 0 & 2000 &			6560 & 0 & 500 \\ 
 &			5807 & 0 & 5000 &			3319 & 0 & 1000 \\ \hline

\end{tabular}
\caption{AWS EKS testing results}
\label{tabla:All-performance-results}
\end{table}

In the first 3 rows (AWS EKS, ECS Fargate and AWS Lambda), the  prototypes do not have a data persistence layer. 

The results of the AWS EKS version show that the sustained load limit over time for this system is between 20 and 30 simultaneous users. 
 For this load, AWS CloudWatch shows processor and memory usage percentages of 60\% and 80\%, respectively. 

The results of the ECS Fargate version are similar to those obtained by the previous version, although slightly worse in terms of sustained load over time. 

It is necessary to remember that the AWS Lambda version is not functional as it does not store modified information except in the memory of the lambda functions themselves. 
For that reason, the latencies obtained by this version are so positive, because access to the data is much faster. 
Still, the auto-scaling capacity of the AWS Lambda function service is clearly appreciated.

In the second scenario, prototypes that integrate the DocumentDB service as a persistence layer are tested.

Apparently, the impact of delegating data storage and access to an external persistence layer, in this case, DocumentDB, is extremely positive in terms of the latencies experienced and the load supported by the application. 
This prototype handled the tests with 10,000 users with a 0\% error rate very solidly. 

The performance results of the ECS Fargate solution with DocumentDB are slightly worse, although still quite positive. The ability of this latest version to handle sudden increases in users is noteworthy. For the performance tests, the CPU and memory usage of the ECS Fargate tasks involved in the system is 0.3\% and 0.79\% respectively, showing that the limiting factor is the memory of the tasks. 

The high latencies experienced in the performance tests by the AWS Lambda with DocumentDB version are due to the time it takes AWS to spin up the necessary Lambda function instances to handle the demand (cold start) and to an architecture in which a request during its life cycle has to go through three services before being answered.
	
In the third scenario, the technologies already seen are combined with another data persistence service, DynamoDB. 
The AWS EKS version with DynamoDB reports good results by passing the tests with 10000 users without major difficulties, which proved challenging for other prototypes. 

The prototype based on ECS Fargate with DynamoDB did not pass the performance tests carried out with 10000 users. 
Once again, the catalog service experienced outages causing a significantly high error rate. The cause of the mentioned outages, once again, turns out to be the memory of the ECS Fargate task. 

The solution based on AWS Lambda with DynamoDB delivers a 0\% error rate. 
The significant increase in latency for the tests with 5000 users is justified by the time it takes for DynamoDB and AWS Lambda to provision itself when there is a surge in users.
	
\section{Discussion} \label{sec:Discussion}

As a summary of the analysis, table~\ref{tabla:performance-testing-summary} shows the average latencies obtained by the best prototypes in performance tests with 10000 user load. The best performance solution was based on AWS EKS with DocumentDB, followed closely by the solution based on AWS EKS with DynamoDB and AWS Lambda with DocumentDB. 

The superiority of AWS EKS service in terms of performance may be due to the provisioning time of AWS Lambda required to handle the load. The prototype based on AWS Lambda with DynamoDB has latencies that are too high due to the provisioning of DynamoDB and the cold-start of AWS Lambda to be among the best. It demonstrates that one disadvantage of automatic scaling is that it is not instantaneous, and its impact on performance must also be taken into account. 
The AWS EKS-based solution offers better latencies because it has provisioned resources and does not require time to scale up. 
\begin{table}[h]
\centering
\begin{tabular}{|c|c|c|}
\hline
\multicolumn{3}{|c|}{\textbf{Performance testing summary}} \\ \hline
\textbf{Prototype} & \textbf{Latency (ms)} & \textbf{Error \%} \\ \hline
AWS EKS + DocumentDB & 61 & 0 \\ \hline
AWS EKS + DynamoDB & 227 & 1.97 \\ \hline
AWS Lambda + DocumentDB & 261 & 0 \\ \hline
\end{tabular}
\caption{performance testing summary}
\label{tabla:performance-testing-summary}
\end{table}

Regarding the monthly cost of each prototype, there is a correlation between Serverless based solutions and those that represent lower costs. The cheapest prototype is based on AWS Lambda with DynamoDB.

\section{Related work} \label{sec:RelatedWork}

Firstly, a survey on Serverless is \parencite{Hassan2021}, which provides a summary of the technology situation and its explanation.
As far as we know, the publication that is more similar to our work is~\parencite{Jin2021}, which is focused on application migration, where the authors migrate 4 complex microservices-based applications to serverless. 
One difference being that work and ours is that it focuses more on the process, 
explaining architectural patterns and giving some tips while our work focuses on the impact of those architectural decisions.
Another study on migrations is \parencite{Yussupov2019}, in which 4 simple Serverless based applications are migrated across 3 cloud providers, 
focusing on aspects of the lock-in problem. Another article similar to ours is \parencite{Lloyd2018}, 
in which Minh Vu et al migrate a system to Serverless using AWS Lambda, 
measuring its performance, scalability, and costs involved. 
Additionally, they experiment on mitigating the cold start of lambda functions. \parencite{Yanaga2017} is a book written by Edson Yanaga focused on migrating monolithic applications to microservices but with a focus on the data and system state perspective.

On the other hand, there are some papers that compare architectures like, 
for example, \parencite{Gos2020}, in which the authors compare a monolithic architecture and 
microservices based architecture in terms of performance, 
strengths, and weaknesses of each. 
Another example is \parencite{Fan2020}, in which Fan et al compare a service-based architecture and a Serverless architecture in terms of latency, cost, scalability, and reliability, concluding that both have their advantages in specific scenarios and that neither is suitable for all scenarios. \parencite{Villamizar2016} compares 3 versions of the same application, one monolithic, 
another based on microservices operated by the cloud customer, and another based on microservices operated by the cloud provider, analyzing the impact on the infrastructure cost, concluding that the microservices architecture can significantly reduce the infrastructure cost. Another publication by BBVA Labs that focuses on serverless costs is \parencite{Alda2022}, where its usage and the usage of EC2 are analyzed in comparison, as well as the parameters that most affect these decisions. 

In addition to this, some publications related to both Serverless and microservices architecture are \parencite{Sbarski2018}, in which Peter Sbarski deeply discusses aspects of serverless such as the most important architectural patterns, authentication and authorization among others, providing several examples of real applications. \parencite{Miell2019}, where Ian Miell exposes the limits to consider when building Lambda functions with AWS Lambda regarding the physical limits of the service. 
\parencite{Jamshidi2018} in which the evolution of the microservices architecture is analyzed, and the challenges of choosing this architecture in the future are reflected upon.

\section{Conclusions and future work} \label{sec:Conclusion}

Before applying the aforementioned comparison criteria to the previous scenarios, it is necessary to highlight that each of these possible architectures and technologies offer advantages and disadvantages that may be beneficial or detrimental for a specific problem. Ultimately, the choice between them is a trade-off with inherent problems and virtues, and such decisions should be made solely by weighing the project's needs and available resources. 

Despite the basic nature of the starting system, the migration processes carried out provide certain ideas as a conclusion. The main idea that stands out is that any transformation process in a system's architecture is not a trivial task and, as such, presents a range of challenges and difficulties to be solved. Other notable issues are that auto-scaling is not free in terms of both cost and performance, and that it is always beneficial to assign a single responsibility to each component of the architecture for its scaling. 

Depending on the initial and target architectures, the necessary process will require more or fewer resources and time, even requiring partial or complete rewriting of the application.

It does not make sense to choose a prototype as the best since, as mentioned before, the appropriate architecture depends on the specific needs of the problem to be solved and other factors that have not been studied for these solutions such as operational complexity, security, etc.

With regards to future work, we are considering to extend this research to compare the prototype implementations using different cloud providers. Another possibility would be to use the tools and scripts that we have used in this study to compare more complex or even real projects.

\subsubsection{Acknowledgements}

This work has been partially funded by a grant provided to the first author 
by the \emph{Cátedra DXC} signed between \emph{DXC Technology} and \emph{University of Oviedo}. 
The authors would like to thank the members of the DXC Assure Team Luciano Campal Vázquez, Pablo Castaño Iglesias, Rubén López Aparicio, Iván Rodríguez Bautista and others, for their support and involvement in the project.

\printbibliography

\end{document}